\documentclass[prl,twocolumn,preprintnumbers,showpacs,longbibliography]{revtex4-1}
\usepackage{amsmath, mathrsfs, amssymb,amsfonts,amsthm,graphicx, epsf, dcolumn, yfonts}
\usepackage[hyperfootnotes=true]{hyperref}
\usepackage{etoolbox}
\usepackage{subfigure}
\usepackage{color}
\usepackage{slashed}
\usepackage{setspace}
\usepackage{cancel}
\usepackage{wasysym}
\usepackage{hyperref}
\pdfoutput=1
	\def\be{\begin{equation}}
		\def\ee{\end{equation}}
	\def\ba{\begin{eqnarray}}
		\def\ea{\end{eqnarray}}

	\def\a{\alpha}
	
	\def\b{\beta}

	\def\d{\delta}

	\def\th{\theta}

	\def\l{\lambda}
	\def\L{\Lambda}

	  \def\cF{{\cal F}}

	  \def\cO{{\cal O}}
	  
	 \def\cT{{\cal T}}

	\newcommand{\prt}[1]{{\left( {#1} \right)}}
	\newcommand{\prtt}[1]{{\left[ {#1} \right]}}

	\def\IR{\relax{\rm I\kern-.18em R}}

	\def\pp{\partial}
	
	\newcommand{\ff}{\frac}

	\def\IR{\relax{\rm I\kern-.18em R}}
	\def\IL{\relax{\rm I\kern-.18em L}}
	
	\def\inv{^{\raise.15ex\hbox{${\scriptscriptstyle -}$}\kern-.05em 1}}

	\def\bea{\begin{eqnarray}}
		\def\eea{\end{eqnarray}}
	\newcommand{\eq}[1]{(\ref{#1})}
	\def\nn{\nonumber}

	\newcommand{\la}[1]{\label{#1}}

	\def\a{\alpha}      
	\def\b{\beta}       
	    
	\def\d{\delta}

	\def\l{\lambda} \def\L{\Lambda}

	\def\th{\theta}

	\definecolor{markcolor2}{rgb}{1,0,0}
	
	\definecolor{markcolor3}{rgb}{0,1,0}

\begin{document}

\title{Holographic Timelike c-function}% Force line breaks with \\
\author{Dimitrios Giataganas${}^{1,2}$}
% \email{dimitrios.giataganas@mail.nsysu.edu.tw}
 \affiliation{${}^{1}$Department of Physics, National Sun Yat-Sen University, Kaohsiung 80424, Taiwan.\\
 ${}^{2}$Physics Division, National Center for Theoretical Sciences, Taipei 10617 Taiwan}%Lines break automatically or can be forced with \\

\begin{abstract}

The integration of high-energy degrees of freedom along the renormalization group (RG) flow in Poincaré-invariant theories is captured by a monotonic c-function. For such theories, holographic c-functions have been constructed using entanglement entropy. However, in theories with broken Lorentz invariance, these constructions generally fail, reflecting both the violation of entanglement RG monotonicity and its limitations in capturing essential features of nonrelativistic RG flows. Since many quantum many-body systems lack Lorentz invariance, it is important to identify a quantity that captures the decrease of effective degrees of freedom along such flows. We show that the recently introduced holographic timelike entanglement entropy naturally gives rise to a new c-function applicable to all such theories, which moreover obeys a geometric upper bound on its rate of change. We further demonstrate the existence of this c-function in theories with Lifshitz and hyperscaling-violating fixed points, showing that, provided the null energy condition, thermodynamic stability and explicit geometric conditions on the flow are satisfied, it obeys the expected RG monotonicity.
 
\end{abstract}

\maketitle

%\tableofcontents

\noindent \textbf{1. Introduction.}
\label{sec::intro}
The renormalization group (RG) flow in quantum field theories reveals a rich structure and serves as a powerful tool for their analysis. One of its most well-known features is the existence of a so-called c-function, which decreases along  RG flows and coincides with the central charge at the conformal fixed points of the theory. This property was first proven in two-dimensional quantum field theories \cite{Zamolodchikov:1986gt} and was later reformulated in terms of entanglement entropy, a quantity that measures how entangled a given spatial subregion is with its surroundings. 

A generalization to even spacetime dimensions was conjectured in \cite{Cardy:1988cwa} and proven in \cite{Komargodski:2011vj} for RG flows between conformal fixed points. For certain conformal field theories (CFTs) in arbitrary dimensions, generalized c-theorems have been established within the holographic framework \cite{Freedman:1999gp}, prompting extensive studies and reformulations using entanglement entropy \cite{Ryu:2006bv}, particularly in the context of holography \cite{Ryu:2006ef, Myers:2010xs, Myers:2010tj, Casini:2011kv, Hung:2011ta, Liu:2012eea, Chu:2019uoh}.

An intuitive understanding of these results comes from interpreting the central charge as a measure of the number of degrees of freedom in the theory. Since RG flow involves integrating out high-energy degrees of freedom \cite{Wilson:1974mb}, a decrease in this measure along the flow is expected. A generalized version of the holographic c-theorem is anticipated to hold in theories with non-relativistic fixed points, which are common in quantum matter and many-body condensed matter systems. Such generalizations have been explored holographically \cite{Cremonini:2013ipa, Chu:2019uoh}, though a universally appropriate c-function for these systems remains elusive. This absence stems from a deeper reason: it is known that RG flow can violate the monotonicity of entanglement entropy under RG transformations \cite{Swingle:2013zla}.

This is not unexpected, since entanglement entropy is defined on a fixed time slice and thus may not capture the full dynamics of non-relativistic RG flows. As a result, it is not surprising that it fails to support the existence of a proper c-function. This observation raises a fundamental question: does there exist a quantity that respects RG flow monotonicity in these broader contexts, and if so, how can it be applied to the physics of quantum matter?

In this work, we introduce a novel c-function based on the recently proposed timelike entanglement entropy in holography \cite{Doi:2023zaf},\footnote{Timelike entanglement entropy and pseudoentropies have been shown to contain valuable information about quantum field theories and their phase transitions, and are the subject of active recent research \cite{Kanda:2023jyi, Afrasiar:2024lsi,  Chu:2023zah,Caputa:2024gve, Afrasiar:2024ldn, Narayan:2022afv, Basak:2023otu, Roychowdhury:2025aye, Chu:2025sjv}.}. This quantity is defined holographically as the area of extremal surfaces-spacelike and timelike-that are anchored to a time subinterval on the boundary \cite{Doi:2023zaf, Afrasiar:2024lsi}. Remarkably, we show that the timelike c-function constructed from this pseudoentropy is monotonic along holographic RG flows. This monotonicity arises non-trivially from the geometric properties of the extremal surfaces involved.

Our proposed c-function maintains the correct monotonic behavior even in physical theories with broken Lorentz or scale invariance. Specifically, we show that the derivative of the timelike c-function along the RG flow can be expressed in a compact way in terms of the bulk geometry of the holographic dual.  As a general test, we verify the most general homogeneous spaces with monomial metric elements, namely with Poincaré invariance, anisotropic Lifshitz scalings, and scale covariant hyperscaling violation all admit a monotonic c-function under RG flow, provided that the null energy conditions (NEC), thermodynamic stability conditions and an effective dimension condition are satisfied. These theories are the most interesting and physically relevant due to the symmetries they possess. Moreover, we emphasize that in the most general fully arbitrary Lorentz violating backgrounds, NEC and thermodynamic stability do not always map one-to-one onto the conditions required for timelike entanglement entropy monotonicity. This is not unexpected; such fully arbitrary backgrounds typically cannot arise as solutions from physically motivated bulk actions. In fact the satisfaction of the timelike c-theorem may suggest that only specific subclasses of these anisotropic geometries, like those we analyzed,  may arise as well-behaved consistent holographic theories.

\noindent \textbf{2. Holographic Setup.} 
%\la{sec::setup}
We consider the homogeneous holographic spacetime in the following coordinate system
\be \la{metric1}
ds^2_{d+1}=-e^{2B(r)} dt^2+e^{2 A(r)}dx^2+dr^2~,
\ee
where $d-1$ is the number of dimensions of the spatial plane spanned by $\vec{x}$ and the boundary of the space-time is taken to be at $r\rightarrow \infty$, without loss of generality.  Static homogeneous spacetimes can always be diagonalized and brought to the form of \eqref{metric1}; therefore, this represents the most generic case.

For a boundary time interval $-\frac{T}{2}\le t\le \frac{T}{2}$, with $x_1=0$ and the remaining coordinates $x_i = \sigma_i$, the bulk surface is parametrized by $t(r)$. The action corresponding to the timelike entanglement entropy is given by
\be\la{s1}
S=2\int dr~ e^{(d-2) A(r)}\sqrt{1-e^{2 B(r)}t'^2}~,
\ee
with the associated equations of motion 
\be \la{t1}
t_{s}=2 \int dr \frac{e^{\L_m-B}}{\sqrt{s  e^{2\L(r)}+  e^{2 \L_m}}}~,\quad s^2:=1~.
\ee
The positive value $s=+1$ corresponds to the spacelike branch of the surface, while $s=-1$ corresponds to the timelike branch. Here, $\Lambda(r):=B(r)+(d-2)A(r)$ and $\Lambda_m:=\Lambda(r_m)$, where $r_m$ is the turning point of the symmetric timelike surface at which $t_{-}'(r)$ diverges. The turning point enters the equation of motion via the integration constant. Substituting \eqref{t1} into \eqref{s1}, we obtain the area in terms of the geometry:
\be \la{s2}
S_{s}=2\int dr\ff{e^{2\L-B}}{\sqrt{e^{2\L}+s e^{2 \L_m}}}~,
\ee
where $S_s$ denotes the total timelike entanglement entropy. As before, the subscript $s=+1$ corresponds to the spacelike part of the bulk surface described by the solution $t_{+}(r)$, contributing the real part of the timelike entanglement entropy $S_{+}$. The integration bounds in this case lie between $r_{\pp}$ and $r_b$, where $r_\pp$ denotes the boundary of the theory and $r_b$ is the deepest point in the bulk reached by the extremal surface. For example, in $AdS$ space in Poincaré coordinates, with the boundary at $r_\pp \rightarrow \infty$, we have $r_b=0$.

The subscript $s=-1$ corresponds to the timelike portion of the surface, described by the solution $t_{-}(r)$, and contributes the imaginary part of the timelike entanglement entropy $S_{-}$. The timelike surface extends from the turning point $r_m$ to $r_b$. The imaginary and real contributions of the timelike surface are denoted by $S_{Im}:=i S_{-}$ and $S_{Re}:=S_{+}$, respectively. Similarly, the time integrals describing these surfaces are denoted by $t_{Re}:=t_{+}$ and $t_{Im}:=t_{-}$, providing a unified notation.

To describe the common class of solutions studied in \cite{Doi:2023zaf, Afrasiar:2024lsi}, the spacelike surface $t_{Re}$ originates from the boundary at $t=T/2$ (with $r=r_\pp \to \infty$), and in theories without horizons or IR walls, it asymptotes to the deep infrared as $r=r_b\to 0$, reaching $t=T_{Re}/2\to \infty$. The right, symmetric branch of the timelike surface $t_{Im}$ begins at the mirror-symmetric extremal point of the surface at $t=0$ (with $r=r_m$), and similarly, in the absence of horizons or IR walls, it asymptotes to the deep IR as $r=r_b\to 0$, with $t=T_{Im}/2\to \infty$. 

\noindent \textbf{3. Holographic Timelike c-function.} 
%\la{sec::cfunction}
The timelike entropy supports a holographic, monotonic c-function for any type of homogeneous theory. To construct a suitable function, we consider the derivatives of $t_s$ and $S_s$, corresponding to the timelike and spacelike surfaces, with respect to the turning point $r_m$ of the timelike surface. We will demonstrate that a unique and naturally motivated c-function exists.

We begin by analyzing the timelike surface. Applying the Leibniz integral rule yields the derivative of the area:
\be \la{ds1}
\ff{1}{2}\ff{\pp S_{im}}{\pp r_m}=\ff{e^{2 \L-B}}{\sqrt{e^{2\L_m}-e^{2\L}}}\bigg|_{  r_m}+
\L_m'e^{2\L_m}\int_{r_m}^{r_b} dr\ff{e ^{2\L-B}}{\prt{e^{2\L_m}-e^{2 \L}}^{\ff{3}{2}}}~.
\ee
Similarly for the time interval we obtain 
\be \la{dt1}
\ff{1}{2}\ff{\pp t_{im}}{\pp r_m}= \ff{e^{ \L-B}}{\sqrt{e^{2\L_m}-e^{2\L}}}\bigg|_{  r_m}+\L_m' e^{\L_m}  \int_{r_m}^{r_b} dr \ff{e^{2 \L-B}}{\prt{e^{2\L_m}-e^{2 \L}}^{\ff{3}{2}}}~.
\ee
By combining equations \eq{ds1} and \eq{dt1}, we obtain the derivative of the timelike part of the area with respect to its time interval:
\be \la{derim}
\ff{\pp S_{Im}}{\pp r_m}= e^{\L_m}\ff{\pp t_{Im}}{\pp r_m}\Rightarrow\ff{\pp S_{Im}}{\pp t_{Im}}=e^{\L_m}~.
\ee
For the spacelike surface, we follow a similar strategy to compute the desired derivative:
\be \la{derre}
\ff{\pp S_{Re}}{\pp t_{Re}}=-e^{\L_m}~.
\ee
Here, the minus sign arises because the boundary of these integrals is independent of $r_m$.

We observe that the derivatives of the spacelike and timelike areas with respect to their corresponding time intervals are equal in absolute value. Motivated by this observation and the holographic definition of the c-function via entanglement entropy \cite{Casini:2006es,Ryu:2006ef,Myers:2012ed,Liu:2012eea,Chu:2019uoh}, we propose a candidate monotonic timelike c-function based on the timelike entanglement entropy. The appropriate function is:
\be \la{cfunction}
c=c_{Re}+c_{Im}~,
\ee
with
\be \la{cf0}
c_{(Im,Re)}:=t_{(Im,Re)}^{d_x}\frac{\pp S_{(Im,Re)}}{\pp t_{(Im,Re)}}
\ee
where $d_x$ is a constant with physical significance, which we will later relate to an effective dimensionality of the observables in the theory.

We can reexpress the c-function \eq{cfunction} in terms of the boundary interval $T$, by absorbing the usual positive normalization constants into the definition:
\be \la{cf0T}
c=T^{d_x} \frac{\pp S}{\pp T}~,
\ee
with $S=S_{Re}+ S_{Im}$ and $T=t_{Im}-t_{Re}$. The expressions of the c-function of \eq{cf0} and of \eq{cf0T} have the same monotonicity along the RG flow and are equivalent.

To investigate the monotonicity of the c-function along the RG flow in generic holographic theories, we analyze the derivatives of its components. It is crucial to note that the time interval can be rewritten as:%dg
\be \la{tx2}
t_{s}=-2 \int dr \ff{e^{\ff{\L}{dx}-B}}{\L'} \cdot F_s~,\quad F_{s}:=\ff{\L' e^{-\ff{\L}{d_x}+\L_m}}{\sqrt{e^{2 \L_m}+s e^{2\L}}}~.%\qquad s^2:=1~,
\ee
We treat the timelike and spacelike parts separately and study each component individually and fix direction of the integral's boundaries appropriately.

\noindent \textbf{3.A. The Timelike Contribution to the c-function.} 
%\la{sec::tilemike1} 
By performing integration by parts on the time interval expression \eq{tx2}, we obtain:
\be \la{im2}
-\frac{t_{Im}}{2}=\frac{e^{\frac{\L}{d_x}-B}}{\L'}\cF_{-} \bigg|_{r_m}^{r_b}-\int_{r_m}^{r_b}\frac{ e^{\frac{\L}{d_x}-B}\cF_{-}}{\L'}\prt{\frac{\L'}{d_x}-\frac{\L''}{\L'}-B'}~.
\ee
The monotonicity of $c_{Im}$ is computed from \eq{cf0} as:
\be \la{cf00}
\frac{\pp c_{Im}}{\pp r_m}=t_{Im}^{d_x-1}e ^{\L_m} \prt{  d_x \ff{\pp t_{Im}}{\pp r_m}+\L_m' t_{Im}}
\ee
We then express the derivative of $t_{Im}$ with respect to itself, making use of the fact that the derivative of the $\cF$ function can conveniently be rewritten in terms of $\cF$. After a lengthy algebraic manipulation, we arrive at:
\bea \nn
\frac{\pp c_{Im}}{\pp r_m}=&&2e^{\L_m} t_{Im}^{dx-1}d_x\L_m'\bigg[-\frac{t_{Im~b}'}{\L_b'}\\
&&+\int_{0}^{\frac{T_{Im}}{2}} dt \frac{1}{\Lambda'}\prt{\frac{\L'}{d_x}-\frac{\L''}{\L'}-B'}\bigg]~, \la{cimtot}
\eea
where we have traded the $r$-integral using the differential of the time-interval length from \eq{t1}, and we emphasize that the derivative of the surface, $t_{Im~b}'$, evaluated at the bulk point $r_b$, appears explicitly in the expression.

\noindent \textbf{3.B. The Spacelike Contribution to the c-function.} 
We now consider the spacelike contribution. Performing integration by parts on the time interval \eq{tx2}, the resulting expression involves $\cF_{+}$. We proceed in a similar way as in the previous subsection. Using the properties of the derivatives of $\cF_{+}$, we relate the derivative of $t_{Re}$ to the function itself. After some algebra, the derivative of the spacelike component of the c-function from \eq{cf0} eventually takes the following form:
\bea\nn
\frac{\pp c_{Re}}{\pp r_m}&=& 2 e^{\L_m} t_{Re}^{d_x-1}d_x\L_m'\bigg[\prt{\frac{t_{Re~b}'}{\L_b'}-\frac{t_{Re~\pp}'}{\L_\pp'}}
\\&&-\int_{\frac{T}{2}}^{\frac{T_{Re}}{2}} dt \frac{1}{\L'}\prt{\frac{\L'}{d_x}-\frac{\L''}{\L'}-B'}~.\la{cretot}
\eea
Here, $T$ is the time interval at the point where the surface originates from the boundary of the theory, and $T_{Re}\to\infty$ denotes the time interval as $r \to r_b$, where the surfaces asymptote to the IR. Notice the crucial detail that the expression in the first line depends on the derivatives $t_{Re}'$ evaluated at $r = r_b$ and $r = r_\pp$. 

%%%%%%%%%%%%%%%%%%%%%%%%%%%%%%%%%%%

\noindent \textbf{3.C. The c-theorem.} 
%\la{sec::cfunction}
%%%%%%%%%%%%%%%%%%%%%%%%%%%%%%%%%%%%
The monotonicity of the c-function \eq{cfunction} is determined by the sum of the derivatives of the spacelike and timelike contributions. The fact that these two contributions are intertwined stems from the asymptotic convergence of the spacelike and timelike surfaces. This becomes evident from the expression for the gradient of the normal vector along both surfaces \cite{Afrasiar:2024ldn}, which reads:
\be 
|\cT|^2=-\frac{e^{2 A\prt{d-1}}}{e^{2 \L}+s e^{2\L_m}}~.
\ee
The smooth merging of the surfaces occurs at a radial position $r_b$. In this limit, $e^{2 \L_b} \to 0$, leading to $|\cT|^2 \sim - s^{-1}$, and the transition from timelike to spacelike surfaces happens smoothly, preserving the correct signature of the normal vector field. This, in turn, imposes the following condition on the derivatives of $t$ at the merging point in the deep IR:
\be\la{condder}
t_{Re~b}'^2\simeq t_{Im~b}'^2~.
\ee
Moreover, in the absence of horizons or IR walls, we have $T_\infty := T_{Re} \simeq T_{Im} \simeq \infty$, reflecting the fact that both surfaces asymptotically approach the IR. Using these properties, and inserting the expressions for the timelike \eq{cimtot} and spacelike \eq{cretot} contributions, the derivative of the total c-function \eq{cfunction} becomes:
\bea\nn
\frac{\pp c}{\pp r_m}&=& 2e^{\L_m} T_{\infty}^{d_x-1}d_x \L_{m}'\bigg[\frac{1}{\L_b'}\prt{t_{Re~b}'-t_{Im~b}'}-\frac{1}{\L_\pp'} t'_{Re~\pp}\\
&&+\int_{0}^{\frac{T}{2}} dt
\frac{1}{\L'}\prt{\frac{\L'}{d_x}-\frac{\L''}{\L'}-B'}\bigg]~.\la{ctot1}
\eea
The first two terms cancel due to the condition \eq{condder}, while the remaining term on the first line vanishes because the surface intersects the boundary orthogonally. This can also be seen from \eq{t1}, since $e^{A_{\pp}} \to \infty$. Furthermore, to obtain \eq{ctot1}, we combined the integrals as follows: $\int_{0}^{\frac{T_\infty}{2}} dt\prt{\ldots} -\int_{\frac{T}{2}}^{\frac{T_\infty}{2}} dt\prt{\ldots}=:\int_{0}^{\frac{T}{2}} dt\prt{\ldots}$,
where the first integral runs along the timelike branch and the second along the spacelike one, so that the right-hand side is shorthand for this signed branch combination. For the adiabatic, slowly varying flows considered here, with the appropriate IR regularization and normalized limit, the signed integral is non-negative when its local integrand is non-negative; see \cite{Giataganas:2025ize}. Eventually, the derivative of the c-function can be compactly written as:
\bea\la{c_final}
\frac{\pp c}{\pp r_m}=2 e^{\L_m} T_{\infty}^{d_x-1}d_x \L_{m}'\int_{0}^{\frac{T}{2}} dt \frac{1}{\L'}\prt{\frac{\L'}{d_x}-\frac{\L''}{\L'}-B'}~.%\la{ctot2}
\eea
This expression is subject to the natural holographic condition at the boundary:
\be \la{nat1}
\frac{t_{Re~\pp}'}{\L_\pp'}= 0~.
\ee
Equation \eq{c_final} is one of the main results of this work. It yields the c-theorem:
\be 
\frac{\pp c}{\pp r_m}\ge0~,
\ee
for a UV boundary located at infinity in the holographic spacetime. This inequality translates into a constraint on the geometric components of the metric, as encoded in \eq{c_final}. The condition takes an averaged form: the integral in \eq{c_final}, multiplied by $d_x \L_m'$, determines the nature of the monotonic behavior of the c-function.

However, a stronger version of this condition can also be imposed: namely, that the integrand in \eq{c_final}, when multiplied by $d_x \Lambda_m'$, maintains a definite sign pointwise along the radial flow. This stricter requirement leads to a fully general formulation of a holographic c-theorem. Moreover, the c-function in \eq{c_final} exhibits an upper bound on its rate of change along the RG flow, 
\be\la{sand0}
0\le\pp_{r_m} c\le e^{\L_m} T_{\infty}^{d_x-1} T \L_m'~. 
\ee
as discussed in App.~A.2, thereby constraining how rapidly effective degrees of freedom can be coarse-grained in consistent holographic theories.

\noindent \textbf{3.D. NEC and Thermodynamic Stability Conditions.}  
%\la{sec::cond} 
Holographic theories, especially those with broken Lorentz invariance, may suffer from instabilities. To avoid such pathologies, we impose two minimal physical requirements: the null energy condition and thermodynamic stability.

The NEC for the metric fields along the holographic RG flow serves as a minimal consistency condition ensuring causal propagation, non-repulsive
gravity, avoiding instabilities and superluminal modes \cite{Wald:1984rg,Buniy:2006xf,Dubovsky:2005xd,Hoyos:2010at}. We obtain two independent conditions which can be written as \cite{Chu:2019uoh,Cremonini:2013ipa} 
\bea \la{nec1}%\nn
&&B''-A''+(B'-A')\prt{B'+(d-1)A'}\ge0~,\\
&&- A(r)'^2+B'(r) A'(r)-A''(r)\ge0~.\la{nec2}
\eea
Thermodynamic stability is analyzed by considering a black hole solution of the metric   \eq{metric1} with the usual blackening factor $f(r)$ satisfying $f(r_h)=0$ at the black hole horizon $r=r_h$. To avoid a conical singularity, the Euclidean compactified time circle must be identified with period set by the Hawking temperature,
\be 
T=\frac{f'(r_h)}{4 \pi}e^{B(r_h)}~.
\ee
The thermal entropy density is proportional to the area of the black hole horizon. By absorbing  constants and positive-definite prefactors, the normalized specific heat, derived from the thermal entropy $S_{th}$, is proportional to  
\be \la{stable}
 \frac{\pp S_{th}}{\pp T}\sim\frac{\prt{d-1} A'(r_h)}{\pp_{r_h}f'(r_h)+f'(r_h) B'(r_h)}\ge 0~.
\ee
The three conditions \eqref{nec1}, \eqref{nec2}, and \eqref{stable}, taken together, guarantee the physical stability and naturalness of the theory.

\noindent \textbf{3.E. Effective Dimensionality.}
\la{sec::effd} 
In this subsection, we consider the theory at a fixed radial scale  $r_f$, and therefore at certain energy scale. The parameter 
$d_x$ can be thought as an effective spatial dimension  of the c-function in the theory we are looking at.  
The integrand of the derivative of the c-function \eq{c_final} vanishes at a fixed radial distance $r_f$ when 
\be \la{dxr1}
d_{x_r}=\frac{\L'(r_f)^2}{\L''(r_f)+B'(r_f) \L'(r_f)}~,
\ee
and therefore $\pp_{r_m} c$ vanishes  when
$d_{x_r}=d_x$ throughout the geometry. 
For theories with $\L''(r_f)\ge 0$,  the NEC \eq{nec1} and \eq{nec2} constrain $d_{x_r}$ to remain positive along the entire RG flow.

The physical meaning of $d_x$  can be understood independently by analyzing the scaling symmetry of the metric. Introducing a new radial coordinate 
$\tilde{r}(r)$ at a particular scale $r_f$, the  following scaling transformation
%\be\la{rescalings}
$t\rightarrow \l^{\tilde{\b}} t~,$ $x\rightarrow \l^{\tilde{a}} x~,$  $\tilde{r}\rightarrow \l^{-\tilde{\d}} \tilde{r}~,$
%\ee
leaves the transformed metric \eq{metric1} 
\be\la{metric2}
ds^2=-\tilde{r}^{2\b} dt^2+\tilde{r}^{2a} dx_i^2+\tilde{r}^{2\d}d\tilde{r}^2,
\ee 
scale covariant, provided $\tilde{a}=\tilde{\d}\prt{\a-1-\d}$ and $\tilde{\b}=\tilde{\d} \prt{\b-1-\d}$. 
The timelike entanglement entropy then scales as 
\be\la{scaleq}
S\sim \tilde{r}^{\a\prt{d-2}+\d+1}\sim t^{-\frac{\a\prt{d-2}+\d+1}{\b-\d-1}}~.
\ee
To ensure that the c-function is dimensionless we obtain from \eq{cf0} and \eq{scaleq} the expression
\be \la{dxresult}
d_{x}=\frac{\b+\a(d-2)}{\b-\d-1}~.
\ee
For the scale covariant theories, or at fixed points with scale covariance the result \eq{dxresult} matches with a direct use of the equation \eq{dxr1} making evident the role of $d_{x_r}$, or $d_x$. This allows us to recast the c-function derivative \eq{c_final} in an even more compact form as
\bea\la{c_final2}
\frac{\pp c}{\pp r_m}= 2 e^{\L_m} T_{\infty}^{d_x-1}\L_{m}'\int_{0}^{\frac{T}{2}} dt\prt{1-\frac{d_x}{d_{x_r}}}~,
\eea
with $d_{x_r}(r)$ given by \eq{dxr1}.

\noindent \textbf{4. Holographic Timelike c-theorem.}
%\la{sec::section3}
\newline
\noindent \textbf{4.A. Timelike c-theorem in Poincare Invariant Theories.} 
Let us consider a conformal theory with $A(r)=B(r)$,
with a bulk geometry described by the metric \eq{metric1}. 
The evolution of the geometry along the RG flow is encoded in the conformal factor $A(r)$. Such geometries are typical solutions of Einstein gravity coupled to a scalar field with an appropriate non-trivial profile along the RG flow, for instance, \cite{Girardello:1998pd,Freedman:1999gp}. The scalar dilaton is dual to a relevant operator $\cO$, and the RG flow can be thought of as being triggered by perturbing the theory at its UV critical point by this operator. In our analysis, the details of the dilaton's profile and potential are not essential.

The effective dimension can be obtained using \eq{dxr1}:
\be 
d_{x_r}=\frac{\prt{d-1}A'^2}{A''+A'^2}
\ee
while the NEC \eq{nec2} gives: $A''\le0$. The integrand of \eq{c_final} becomes up to positive factors,
\be\la{poincare1}
\frac{\pp c}{\pp r_m}\sim A_m'\int_0^\frac{T}{2} dt \prt{1-\frac{d_x}{d_{x_r}}}~,
\ee
which implies that, as long as the norm of the effective dimension $d_{x_r}$ decreases along the RG flow—or alternatively, as long as a constant $d_x$ is equal or smaller than the minimum effective dimension, the c-theorem automatically holds.

To prove this statement, note that the NEC implies $A'$ is a decreasing function along the RG flow.
Since we have placed the boundary of the theory at $r\rightarrow \infty$, we have $A_{UV}\ge 0$ and $A_{\min}'=A_{UV}'$. Therefore, it suffices to show that the minimum value of $A'$ at the boundary is positive to ensure the monotonicity of the c-function \eq{poincare1}. Indeed, this is the case: since $A_{UV}\sim A(r)\rightarrow \infty$ at the boundary, we have $A'_{\min}\ge0$, which then implies $A'\ge0$ along the entire RG flow.

As an explicit example consider the case where at UV, $d_x=d-1$, which is equal to the lower bound of a positive $d_{x_r}$,  by the saturation of the NEC. The monotonicity of the timelike holographic c-function is then always
\be
\frac{\pp c}{\pp r_m}\sim A_m'\int_0^{T/2}dt\frac{\prt{-A''}}{A'^2}\ge 0~,
\ee
which is ensured by the NEC.

Hence, the timelike holographic c-function for a Poincaré invariant theory is monotonic as long as the NEC are satisfied and the effective dimension decreases. This case also serves as a consistency check, confirming that the timelike c-function is an appropriate choice for Poincaré invariant theories. These monotonicity statements apply within the adiabatic class. Beyond this class, the NEC alone may not guarantee monotonicity \cite{Giataganas:2025ize}.

\noindent \textbf{4.B. Timelike c-theorem in Non-Relativistic Theories.}
\la{sec::b30}\newline
\noindent \textbf{4.B.1. Lifshitz Theories.}
\la{sec::b31}
Consider a Lifshitz theory where $A=r$ and $B=z r$, where $z$ is  the Lifshitz exponent, a parameter characterizing Lorentz violation. Applying the analysis of section 3.E we find that  
the metric in Poincare coordinates has exponents $\a=1$, $\b=z$ and $\d=-1$. 
The effective dimension here is given by \eq{dxr1}, or equivalently by \eq{dxresult} and becomes
\be \la{dxlif}
d_{x}=1+\frac{d-2}{z}~,
\ee
while the NEC reduces to the simple inequality $z\ge 1$. When $z=1$ we recover the usual AdS spacetime and dual to the conformal  theory. The NEC constrains $d_{x_r}>0$ and $\L'=z+d-2>0$. 
For the trivial, scale-invariant Lifshitz RG flow, the holographic timelike c-function \eq{c_final}, with $d_x=d_{x_r}$ given by \eq{dxlif}, yields a vanishing derivative: $\pp_{r_m} c=0$, as expected. For a non-trivial Lifshitz RG flow, the validity of the c-theorem \eq{c_final2} is guaranteed as long as $d_x\le d_{x_r}$, since already $\L'>0$ by the NEC. 

Let us elaborate further by considering a theory where the Lifshitz exponent $z_r(r)$ varies very slowly along the RG flow with fixed points characterized by $z_{UV}$ and $z_{IR}$. $d_x$ is constant and is determined at a fixed point, while $d_{x_r}$ is the effective dimension at a given radial distance. Then  \eq{c_final2} approximately gives 
\be \la{lif11}
\frac{\pp c }{\pp r_m}\sim \L_{m}'\int_{0}^{\frac{T}{2}}dt \frac{1}{z\L'} \prt{z-z_{r}} \ge0 ~,
\ee
The NEC  enforces $\L'\ge0 $ and $z_r(r)$ denotes the slowly varying Lifshitz exponent at $r$. Therefore, the c-theorem reduces to the simple condition: $z\ge z_r$. 

For example, the validity of the $c$-theorem can be demonstrated in theories with conformal UV fixed points, $z_{\rm UV}=1$, that satisfy the NEC, so that $z_{\min}=z_{\rm UV}$, and have a slowly decreasing $z_r(r)$. Taking $d_x := d_{xIR}$ and thus $z := z_{IR}=z_{max}$ ensures $z \geq z_r$ and the c-theorem \eq{lif11} holds.

In summary, the NEC together with the universal condition $d_x\le d_{x_r}$ automatically ensures the validity of the c-theorem for Lifshitz RG flows.

\noindent \textbf{4.B.2. Hyperscaling Violation Theories.}
\la{sec::b32}
Next, let us consider a more intricate case: a hyperscaling violation theory characterized by spatial homogeneity and scale covariance, as a stringent test for the c-theorem. Here we have
\be 
\b=z-\frac{\th}{d-1}~,\quad \a=1-\frac{\th}{d-1}~,\quad \d=\a-2~,
\ee
where $z$ and $\theta$ are the dynamical Lifshitz exponent and hyperscaling violation exponent, respectively. 

The effective dimension \eq{dxresult} of the theory is
\be \la{dxhysca}
d_{x}=1+\frac{d-\th-2}{z}~,
\ee
which, when compared with \eq{dxlif}, clearly reflects the effect of the hyperscaling violation exponent: the modes propagate effectively in $d-\theta$ spatial dimensions rather than in $d$, leading to consequences analogous to those seen in the thermodynamics and thermal entropy of the theory.

The c-function evolves along the RG flow as
\bea\la{c_hysca}
\frac{\pp c}{\pp r_m}\sim  \L_m'\int_{0}^{\frac{T}{2}} \prt{1-\frac{d_x}{d_{x_r}}} dt ~,
\eea
where $\L_m'=(d-1)\prt{\th-d-z+2}/\prt{\th r_m}$. The NEC \eq{nec1}, \eq{nec2} and stability conditions \eqref{stable} impose $d_{x_r}\ge0$ for any physical hyperscaling violation theory, but now they do not uniquely fix the sign of $\L'$. We will show that as long as the natural conditions are satisfied, the c-theorem holds for $d_x \le d_{x_r}$, as in our previous RG flows.

To establish the existence of the $c$-theorem, we carefully study each parameter subregime in our coordinate system. There are three consistent subregimes to consider:
A) For $\a> 1$, we have the range $r\in[0,\infty)$ in the coordinate system \eq{metric1}, with a boundary $\pp$ at $\infty$. Once the NEC and the thermodynamic stability conditions are satisfied, the function $\L_{m}'$ is automatically positive and therefore, from \eq{c_hysca}, the derivative $\pp_{r_m}c\ge0$. The c-theorem holds for the range of parameters $\a> 1$. 

For the range of parameters described by B)  $\a<0$ the boundary in the coordinate system \eq{metric1} is again at infinity. The NEC and the thermodynamic stability then impose $\L_{m}'\ge0$ and therefore $\pp_{r_m}c\ge 0$. The c-theorem is satisfied for this case as well.

For subregime C), where $0<\a<1$ the boundary is at the minimum radial value, $r=0$. The NEC and the thermodynamic stability now conveniently impose the opposite conditions compared to previous cases: $\L_{m}'\le0$, and hence, from \eq{c_hysca}: $\pp_{r_m}c\le 0$.  The c-theorem is again satisfied.   

This is a particularly elegant derivation: the derivative of the c-function is naturally constrained by the NEC and thermodynamic stability conditions to change sign depending on the location of the UV fixed point along the radial direction. It takes negative values, as expected, when the UV fixed point lies at the minimum of the radial coordinate-precisely the condition under which the c-theorem holds. Everything fits together seamlessly to ensure the validity of the c-theorem. \newline
\noindent \textbf{4.B.3. Anisotropic Hyperscaling-Violating Theories.}
In \cite{Giataganas:2025ize}, theories in which rotational symmetry in the spatial plane is further broken were considered. This breaking of spatial rotational invariance leads to direction-dependent effective dimensionalities, generalizing \eq{dxr1}, and consequently to a family
of directional $c$-functions. It was shown in \cite{Giataganas:2025ize} that these anisotropic c-functions remain monotonic along the RG flow even in such highly nontrivial backgrounds, with their evolution governed by a general expression analogous to \eq{c_final}. This provides one of the most stringent consistency tests of monotonicity.

\noindent \textbf{5. Discussion.}

We have constructed a c-function based on holographic timelike entanglement entropy that is applicable to non-relativistic theories. We demonstrate that renormalization group flows without Lorentz invariance obey a timelike entanglement monotonicity, thereby establishing a direct connection between RG irreversibility and timelike entanglement. This constitutes a systematic improvement over previous attempts to construct c-functions from spacelike entanglement entropy, whose monotonicity is known to fail once Lorentz symmetry is broken.

Many quantum many-body systems  are inherently non-Lorentz-invariant, motivating the study of their holographic RG flows as a powerful framework for probing strongly coupled dynamics and phase transitions. Our results open new foundational directions for analyzing such RG flows holographically and strongly suggest that, in the absence of Lorentz symmetry, a field-theoretic c-function should be constructed from timelike entanglement measures rather than from spacelike entanglement entropy.   Our work also suggests a more general lesson: when space and time scale differently, RG irreversibility must be probed through observables sensitive to the temporal direction.

Moreover, we identify an upper bound on the rate of change of the timelike c-function along the RG flow, constraining how rapidly effective degrees of freedom can be coarse-grained in consistent holographic theories. This bound reflects the locality and internal consistency of the RG process, and provides a quantitative measure of how fast RG irreversibility can proceed under controlled physical conditions.

\la{sec::con}

\noindent \textbf{Acknowledgment.} We would like to thank T. Takayanagi for valuable correspondence. The research work of DG is supported by the National Science and Technology Council (NSTC) of Taiwan with the Young Scholar Columbus Fellowship grant 114-2636-M-110-004.

%\section{Supplementary material}
\noindent ~\centerline{\textbf{Appendices}}~\newline
\noindent \textbf{A.1. The derivative of the c-function.} 
In this section, we provide additional technical details on the derivation of the derivative of the c-function \eq{c_final}. The timelike interval length, written as in \eq{tx2}, is 
\be 
t_{Im}=-2 \int_{r_m}^{r_b}dr \frac{e^{\frac{\L}{d_x}-B}}{\L'}F_{-} ~,
\ee
and can be rewritten in the form \eq{im2}. After some algebra, its derivative takes the form
\bea\nn
&&-\ff{1}{2} \frac{\pp t_{im}}{\pp r_m}=\frac{\pp_{r_m}\cF_{b-}}{\L_b'}e^{\frac{\L_b}{d_x}-B_b}-\frac{\pp_{r_m}\cF_{m-}}{\L_m'}e^{\frac{\L_m}{d_x}-B_m}\\\la{dera0}
&&-\int_{r_m}^{r_b} dr \frac{e^{\frac{\L}{d_x}-B}\pp_{r_m}\cF_{-} }{\L'} \prt{\frac{\L'}{d_x}-\frac{\L''}{\L'}-B'}~.\la{der2}
\eea
To further process the expression above, we note that the derivatives of the 
$\cF$ function relate to the function itself as follows:
\be\la{dera1}
\frac{\pp_{r_m}\cF_{m-}}{\cF_{m-}}=-\frac{\L_m'}{d_x}~,~~ \pp_{r_m}\cF_{-}+\ff{\L_m'\cF_{-} }{d_x}=-\ff{\L_m' e^{\L_m-\frac{\L}{d_x}}}{\sqrt{e^{2 \L_m}-e^{2\L}} }~.
\ee
In the first expression $\cF_{m-}$  is evaluated before taking the derivative. Using these relations and referring back to  \eq{im2},  we observe that the derivative of   $t_{Im}$ depends on $t_{Im}$ itself. Consequently \eq{dera0} can be rewritten as:
\bea\nn
&&-\ff{1}{2}\prt{ \frac{\pp t_{im}}{\pp r_m}+\frac{\L_m'}{ d_x}t_{Im}}= \frac{e^{\frac{\L_b}{d_x}-B_b}}{\L_b'}  \prt{ \pp_{r_m}\cF_{b-}+\frac{\L_m'\cF_{b-}}{d_x}}\\
&&
+\L_m' e^{\L_m} \int_{r_m}^{r_b} dr\frac{ e^{-B}}{\L'\sqrt{e^{2 \L_m}-e^{2\L}}} \prt{\frac{\L'}{d_x}-\frac{\L''}{\L'}-B'}~.\la{der4}
\eea
The first line simplifies with the help of the second identity in \eq{dera1}. The above expression is essentially proportional to the derivative of the timelike component $c_{Im}$ as can be seen from \eq{cf00}. From this, we obtain a compact expression: 
\bea \nn
&&\frac{\pp c_{Im}}{\pp r_m}= -2t_{Im}^{dx-1} \L_m' e^{2\L_m} d_x\bigg[-\ff{ e^{-B_b}}{\L_b'\sqrt{e^{2 \L_m}-e^{2\L_b} }}\\
&&+ \int_{r_m}^{r_b} dr \frac{e^{-B}}{\L'\sqrt{e^{2 \L_m}-e^{2\L}}}\prt{\frac{\L'}{d_x}-\frac{\L''}{\L'}-B'}\bigg]~. \la{der5}
\eea
Note that in the first term the derivative $t_{Im,b}'$ \eq{t1} of the timelike surface at the deepest point in the bulk $r_b$ appears, and in the integral term we encounter the differential of the time-interval
\be \la{dtimm}
dt_{Im}=-\frac{e^{\L_m-B}}{\sqrt{e^{2\L_m}-e^{2 \L}}} dr~,
\ee
which, when substituted into \eq{der5}, yields the final expression \eq{cimtot}. 

Next, we analyze the spacelike part of the 
c-function, where some differences in the computation arise. The time interval is given by%dg
\be 
t_{Re}=-2 \int_{r_\pp}^{r_b}dr \frac{e^{\frac{\L}{d_x}-B}}{\L'}F_+ ~,
\ee
and can be rewritten as
\bea \nn
-\frac{t_{Re}}{2}=\frac{e^{\frac{\L}{d_x}-B} }{\L'}\cF_{+}\bigg|_{r_\pp}^{r_b}
-\int_{r_\pp}^{r_b}dr \frac{e^{\frac{\L}{d_x}-B}\cF_+}{\L'}\prt{\frac{\L'}{d_x}-\frac{\L''}{\L'}-B'} ~.\la{tre}
\eea
Taking the derivative of $t_{Re}$ yields
\bea \nn
-\frac{1}{2}\frac{\pp t_{Re}}{\pp r_m}=&&\frac{e^{\frac{\L_b}{d_x}-B_b}}{\L_b'}\pp_{r_m}\cF_{b+}-\frac{e^{\frac{\L_\pp}{d_x}-B_\pp}}{\L_\pp'}\pp_{r_m}\cF_{\pp+}\\&&\nn
-\int_{r_\pp}^{r_b} dr 
\frac{e^{\frac{\L}{d_x}-B}}{\L'}\pp_{r_m}\cF_{+}\prt{\frac{\L'}{d_x}-\frac{\L''}{\L'}-B'}~.
\eea
Using  
\be 
\pp_{r_m}\cF_+=-\frac{\L_m'}{\sqrt{e^{2 \L}+e^{2 \L_m}}}e^{-\frac{\L}{d_x}+\L_m}-\frac{\cF_+}{d_x}\L_m'~,\la{derivfp}
\ee
we observe that the derivative of $t_{Re}$  can be expressed in terms of the time-interval itself, and the above expressions simplify conveniently as
\bea \nn
&&-\frac{1}{2}\prt{\frac{\pp t_{Re}}{\pp r_m}+\frac{\L_m'}{  d_x}t_{Re}}=
\\\nn 
&&-\L_m'e^{\L_m}\bigg[\frac{e^{-B_b} }{\L_b'\sqrt{e^{2 \L_b}+e^{2\L_m}}}-\frac{ e^{-B_\pp}}{\L_\pp'\sqrt{e^{2 \L_\pp}+e^{2\L_m}}}\bigg]\\
&&+\L_m'\int_{r_\pp}^{r_b} dr\frac{e^{\L_m-B}}{\L'\sqrt{e^{2 \L}+e^{2\L_m}}}\prt{\frac{\L'}{d_x}-\frac{\L''}{\L'}-B'}~.\la{tre4}
\eea
Nicely, the first two terms are proportional to  $t_{Re~b}'$ and $t_{Re~\pp}'$,  the derivatives of the spacelike surface at the deepest point in the bulk and at the boundary, respectively. Furthermore, the integrand's differential is naturally expressed in terms of the time differential:
\be  \la{dtre}
dt_{Re}=-\frac{e^{\L_m-B}}{\sqrt{e^{2\L}+e^{2 \L_m}}} dr~.
\ee
Equation \eq{tre4} essentially gives $\pp_{r_m} c_{Re}$, since from \eq{cf0}, we then obtain
\be 
\frac{\pp c_{Re}}{\pp r_m}=-t_{Re}^{d_x-1}e ^{\L_m} d_x\prt{ \ff{\pp t_{Re}}{\pp r_m}+\frac{\L_m' t_{Re}}{d_x}}.
\ee
A straightforward substitution of \eq{tre4} completes the derivation of \eq{cretot}. We highlight how elegantly all the geometric properties of the surface combine to yield a compact and easily accessible final expression.

\noindent \textbf{A.2. Bounds on the Rate of the c-Function along the RG Flow.} 
The boundary time interval is given by $T=t_{Im}-t_{Re}$. We can compute the derivative
\be
\frac{\pp c}{\pp T}=\frac{\pp c}{\pp {r_m}} \frac{\pp r_m}{\pp {T}}~,
\ee
where the first factor on the right-hand side is given by the expression \eq{c_final}, which encodes the monotonicity conditions we have analyzed so far. To compute the second factor, we use equations \eq{der4}, \eq{dtimm} for $t_{Im}$ contribution, and   \eq{tre4} and \eq{dtre} for the $t_{Re}$ contribution. This leads us to the compact expression:   
\be 
\frac{\pp c}{\pp T}=\frac{d_x ~\pp_{r_m} c}{\prt{-T \L_m'+ e^{-\L_m}T_{\infty}^{1-d_x}\pp_{r_m} c}}~.
\ee
The c-theorem gives $\pp_{r_m}c\ge 0$, and  $\pp_T c\le0$ since  probing larger intervals corresponds to lower energies, along which the $c$-function decreases. Then depending on the sign of $d_x$, $\pp_{r_m} c$ is bounded from above or below with respect to $T \L_m'$. For example, for positive $d_x$ the c-theorem implies
\be\la{sand}
0\le\pp_{r_m} c\le e^{\L_m} T_{\infty}^{d_x-1} T \L_m'~. 
\ee
The c-theorem sets a lower bound and an upper bound, not necessarily independent of each other, on the derivative of the c-function for RG flows. 

To be more precise we rewrite the expression in terms of the metric elements as
\be 
\frac{\pp c}{\pp T}= T_{\infty}^{d_x-1} e^{\L_m} d_x \prtt{1-\frac{T}{2 d_x  \int_0^\frac{T}{2} dt \frac{1}{\L'}\prt{\frac{\L'}{d_x}-\frac{\L''}{\L'}-B'}}}^{-1}
\ee
and therefore, for instance, for positive $d_x$ the c-theorem imposes
\be \la{up1}
\frac{T}{2d_x  \int_0^\frac{T}{2} dt \frac{1}{\L'}\prt{\frac{\L'}{d_x}-\frac{\L''}{\L'}-B'}}\ge1~,
\ee
where the value of the integral is constrained with respect to the rescaled time interval $T/d_x$.  

One may wonder whether this is a sensible bound and whether it is consistent  with the c-theorem as we have formulated it so far. In fact, it does. To see this, let us rewrite \eq{up1} for positive $d_x$, in terms of $d_{x_r}$ \eq{dxr1}. 
\be \la{up1b}
\frac{T}{2  \int_0^\frac{T}{2} dt \prt{1-\frac{d_x}{d_{x_r}}}}\ge1~,
\ee
Once the $c$-theorem condition $d_x \le d_{x_r}$ is imposed, the first inequality in \eq{sand} is satisfied. Moreover, the integral appearing in the above expression is positive  and \eq{up1b} can be rewritten as $\int_0^\frac{T}{2} dt f(t)\le \frac{T}{2}.$   Equations \eq{c_final} and \eq{dtrm} show that this inequality is equivalent to the interval-orientation assumption $\partial T/\partial r_m\le0$ stated above, so no constraint beyond it arises. In other words, this serves as an additional nontrivial consistency check: the conditions we imposed for the c-theorem by studying $\partial_{r_m} c$ are sufficient to ensure the correct monotonicity along the RG flow when examined through $\partial_T c$.

\noindent \textbf{A.3. Equivalence between the c-function expressions.}
In this section we show that the two expressions of the  c-function  \eq{cf0} and \eq{cf0T} are equivalent. Let us study the monotonicity of \eqref{cf0T} by taking the derivative with respect to $r_m$ to obtain
\be \la{cf0T1}
\frac{\pp c}{\pp r_m}=2T^{d_x-1}e^{\L_m}\prtt{d_x \frac{\pp T}{\pp r_m}+ T\L_m'}~.
\ee
We can compute, using \eq{der4} and \eq{tre4}, the derivative
\be\la{dtrm}
\frac{\pp T}{\pp r_m}=2\L_m'\prtt{-\frac{T}{2 d_x} +
\int_{0}^{\frac{T}{2}} dt \frac{1}{\L'}\prt{\frac{\L'}{d_x}-\frac{\L''}{\L'}-B'}}~.
\ee
%dg fixed 1/\Lamda
Substituting \eq{dtrm} into \eq{cf0T1}, we obtain
\bea
\frac{\pp c}{\pp r_m}=4 e^{\L_m} T^{d_x-1}d_x \L_{m}'\int_{0}^{\frac{T}{2}} dt \frac{1}{\L'}\prt{\frac{\L'}{d_x}-\frac{\L''}{\L'}-B'}~.%\la{ctot2}
\eea
The right-hand side is equal, up to a positive prefactor, to \eqref{c_final}, thus establishing the equivalence of the monotonicity conditions for the two expressions \eqref{cf0} and \eqref{cf0T}.
\noindent

\bibliography{timelike}

\end{document}